\documentclass[aps,prb,twocolumn,superscriptaddress,showpacs]{revtex4}

\usepackage{graphicx, bm, amsmath}

\begin{document}


\title{Phonon Dispersion Relations in PrBa$_2$Cu$_3$O$_{6+x}$ ($x \approx 0.2$)}



\author{C.\,H. Gardiner}
\email[Electronic address before May 1, 2004: ]{carol.gardiner@npl.co.uk}
\email[Electronic address after May 1, 2004: ]{carol.webster@npl.co.uk}
\affiliation{National Physical Laboratory, Queens Road, Teddington, Middlesex, TW11 0LW, UK}

\author{A.\,T. Boothroyd}
\affiliation{Clarendon Laboratory, University of Oxford, Parks Road, Oxford, OX1 3PU, UK}

\author{B.\,H. Larsen}
\affiliation{NKT Research and Innovation, Blokken 84, 3460 Birker{\o}d, Denmark}

\author{W. Reichardt}
\affiliation{Forschungszentrum Karlsruhe, IFP, Postfach 3640, D-76021 Karlsruhe, Germany}

\author{\mbox{A.\,A. Zhokhov}}
\affiliation{Russian Academy of Sciences, Institute of Solid State Physics, Chernogolovka 14232, Russia}

\author{N.\,H. Andersen}
\affiliation{Ris\o\ National Laboratory, Frederiksborgvej 399, P.O. 49, DK-4000 Roskilde, Denmark}

\author{S.\,J.\,S. Lister}
\affiliation{Oxford Magnet Technology Ltd, Wharf Road, Eynsham, Witney, Oxfordshire, OX29 4BP, UK}

\author{A.\,R. Wildes}
\affiliation{Institut Laue-Langevin, Bo{\^i}te Postale 156, F-38042, Grenoble C{\'e}dex 9, France}



\date{\today}

\begin{abstract}
We report measurements of the phonon dispersion relations in
non-superconducting, oxygen-deficient PrBa$_2$Cu$_3$O$_{6+x}$ ($x
\approx 0.2$) by inelastic neutron scattering.  The data are
compared with a model of the lattice dynamics based on a common
interatomic potential.  Good agreement is achieved for all but two
phonon branches, which are significantly softer than predicted.
These modes are found to arise predominantly from motion of the
oxygen ions in the CuO$_2$ planes.  Analogous modes in
YBa$_2$Cu$_3$O$_6$ are well described by the common interatomic
potential model.
\end{abstract}



\maketitle


\section{Introduction}

The suppression of superconductivity by praseodymium in
PrBa$_2$Cu$_3$O$_{6+x}$ has for some years presented an
outstanding problem in the field of cuprate superconductivity.
\cite{Radousky, Boothroyd:PrBCOReview}  Although superconductivity
at $T_{\rm c} \approx 90$\,K is exhibited by the majority of
compounds with the composition $R$Ba$_2$Cu$_3$O$_7$, where $R$ is
a rare-earth, PrBa$_2$Cu$_3$O$_{6+x}$ combines antiferromagnetic
ordering with semiconducting resistivity across the whole of the
known phase diagram ($0 < x < 1$).

An influential model of the electronic structure of
PrBa$_2$Cu$_3$O$_{6+x}$ \cite{Fehrenbacher} suggests that the
absence of superconductivity is caused by localisation of holes
due to hybridisation of the Pr 4$f$ and O 2$p$ orbitals.  This is
supported by the observation of an enhanced ordering temperature
$T_{\rm Pr}$ of the Pr sublattice, which is an order of magnitude
larger than the ordering temperature $T_R$ of the rare-earth
sublattice in other $R$Ba$_2$Cu$_3$O$_7$ compounds.

Neutron scattering studies of the magnetic structure
\cite{Lister:2001} and dispersive magnetic excitations
\cite{Boothroyd:1997, Gardiner:2002:PrBCO} have probed the
strengths of the different magnetic couplings in
PrBa$_2$Cu$_3$O$_{6+x}$, and indicate that the high value of
$T_{\rm Pr}$ is partly due to an enhanced Pr--O--Pr superexchange
and partly due to the presence of a significant coupling between
the Cu and Pr sublattices.  Both of these effects could be the
result of hybridisation between Pr 4$f$ electrons and the CuO$_2$
planes.

Hybridisation is expected to affect the phonon dispersion
relations through the Pr--O bonds.  In this paper we present
measurements of the phonon dispersion relations, performed by
inelastic neutron scattering on an oxygen-deficient single crystal
of PrBa$_2$Cu$_3$O$_{6+x}$ ($x \approx 0.2$).  The data are
compared with a model of the lattice dynamics based on a common
interatomic potential, which was originally developed for
YBa$_2$Cu$_3$O$_6$ and later modified for PrBa$_2$Cu$_3$O$_6$.
\cite{Chaplot}

\section{Experimental details}

The single crystal sample of PrBa$_2$Cu$_3$O$_{6+x}$ was prepared
by top seeding a flux and had a mass of $\sim$ 2\,g.  It was
reduced at 700$^{\circ}$C in a flow of 99.998\% argon for 100
hours and quenched to room temperature resulting in an estimated
oxygen content of $x \approx 0.2$.  Using x-ray Laue diffraction
it was aligned with the [1$\bar{1}$0] direction vertical.  It was
then glued onto an aluminium mount using G.E. varnish and held
securely in place with aluminium wire. Preliminary neutron
diffraction measurements showed the crystal mosaic to be $\sim
1^{\circ}$.

\begin{figure*}[!ht]
\begin{center}
\includegraphics{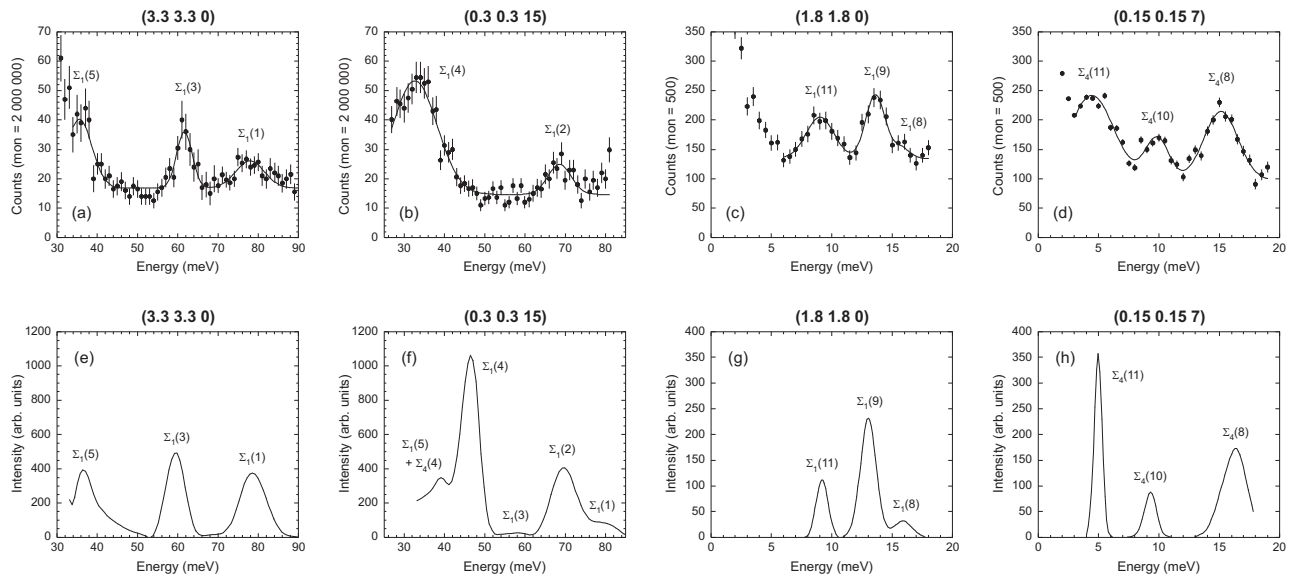}
\caption{Plots (a), (b), (c) and (d) show typical energy scans.
The points are the raw neutron scattering data (averaged over
several runs with different, overlapping energy ranges), and the
lines are Gaussian fits. The value of ${\bf Q}$ is shown at the
top of each plot.  In plots (a) and (b), from IN1, a monitor count
of $2 \times 10^6$ corresponds to a counting time of 2--5 minutes
per point (depending on energy), while in plots (c) and (d), from
IN22, a monitor count of 500 corresponds to $\sim$ 2 minutes per
point. Plots (e), (f), (g) and (h) show simulations of the
energy-scans, produced from the common interatomic potential model
described in the text, which facilitate the assignment of symmetry
labels to the different phonon branches.}
\label{fig:LOandTOscans}
\end{center}
\end{figure*}

The neutron scattering experiments were performed on the IN1 and
IN22 triple-axis spectrometers at the Institut Laue-Langevin
research reactor.  The configuration of IN1 was as follows: flat
copper (200) monochromator, pyrolytic graphite (002) analyser with
horizontal and slight vertical focussing, no collimation before
the monochromator, 60$^{\prime}$ collimator between the
monochromator and sample, and pyrolytic graphite filter between
the sample and analyser to reduce contamination of the beam by
second and higher order reflections from the monochromator.  The
spectrometer was operated in the constant $k_{\rm f}$
configuration with a final energy of 34.8\,meV.  Low efficiency
beam monitors, placed before the monochromator and between the
filter and the analyser, were used to gauge the incident beam
intensity and to check for accidental Bragg scattering.  The
crystal was mounted inside a displex refrigerator and aligned such
that the [110] and [001] directions lay within the scattering
plane.  All measurements were performed at $T$ = 12\,K.  The
configuration of IN22 was identical except for the use of a
vertically curved pyrolytic graphite (002) monochromator, no
collimators and a fixed final energy of 14.7\,meV.  A low
efficiency beam monitor placed before the monochromator was used
to gauge the incident beam intensity, but there was no monitor
between the filter and the analyser to check for accidental Bragg
scattering. The crystal was mounted inside a helium cryostat and
aligned such that the [110] and [001] directions lay within the
scattering plane.  Most of the measurements were performed at a
temperature of 1.6\,K.

\section{Measurements}

The phonon dispersion relations in PrBa$_2$Cu$_3$O$_{6.2}$ were
measured by performing a series of constant ${\bf Q}$ energy scans
at positions across the Brillouin zone (${\bf Q}$ is the neutron
scattering vector).  On IN1, scans were performed over an energy
range $\hbar\omega$ = 30--90\,meV with \mbox{${\bf Q}$ =
($3\!\pm\!h$, $3\!\pm\!h$, 0)} to excite predominantly
longitudinal modes and \mbox{${\bf Q}$ = ($h$, $h$, 15)} to excite
predominantly transverse modes.  On IN22, scans were performed
over an energy range $\hbar\omega$ = 0--24\,meV with \mbox{${\bf
Q}$ = ($2\!-\!h$, $2\!-\!h$, 0)}, \mbox{(0, 0, $7\!+\!l$)} to
excite predominantly longitudinal modes and \mbox{${\bf Q}$ =
($h$, $h$, 7)}, \mbox{($h$, $h$, 8)}, \mbox{(2, 2, $l$)} to excite
predominantly transverse modes. Large values of ${\bf Q}$ were
chosen to maximise the phonon scattering cross section (which is
proportional to $Q^2$), while minimising the cross section for
magnetic crystal field excitations (which scales with the magnetic
form factor, falling off at high $Q$). \cite{Squires}

The scattering intensity of a phonon mode is determined by its
dynamical structure factor, a quantity that varies within each
Brillouin zone and also between different zones.  This means that
some modes have a measureable intensity only at certain points
within each Brillouin zone.  In such cases, or when a scan was
contaminated by accidental Bragg scattering, energy scans were
performed in neighbouring ($\pm h$) Brillouin zones to gain as
many points on the dispersion curves as possible.

Some typical scans taken on IN1 and IN22 are shown in Figs.\
\ref{fig:LOandTOscans} (a)--(d). The peaks are quite broad because
of the horizontal focussing of the analysers.  There is little
evidence of magnetic scattering due to crystal field excitations,
which are known to exist in PrBa$_2$Cu$_3$O$_6$ at energies of
61.5, 65.2, 76.0 and 84.7\,meV, \cite{Hilscher}, but it is
possible that a small amount of magnetic scattering might
contribute to the broadness of the observed phonon branches in the
energy range 70--90\,meV.

\begin{figure}[!ht]
\begin{center}
\includegraphics{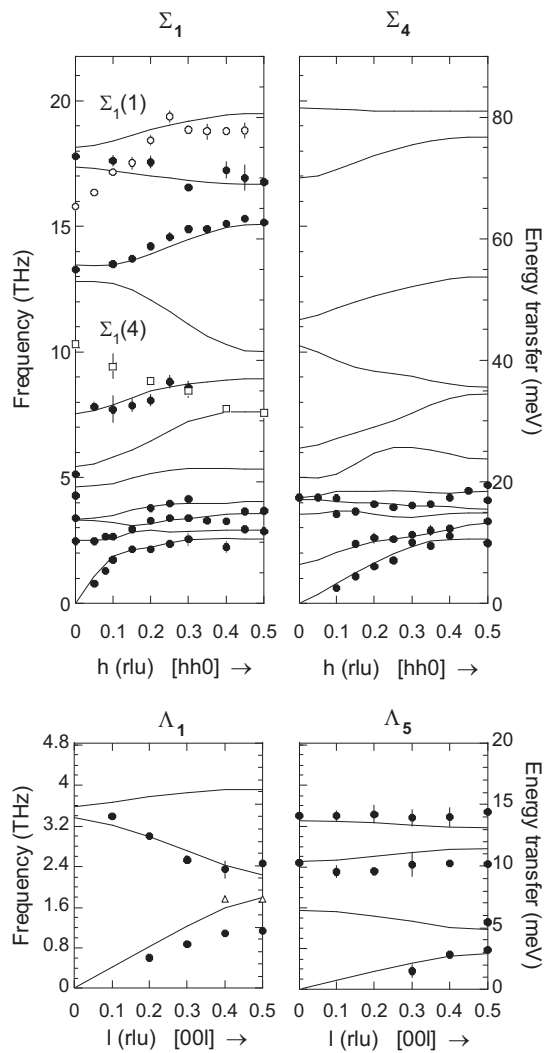}
\caption{Phonon dispersion curves in PrBa$_2$Cu$_3$O$_{6.2}$.  The
points are the fitted peak centres obtained from the constant
${\bf Q}$ energy scans.  The lines are the dispersion curves
predicted by the common interatomic potential model, adapted from
YBa$_2$Cu$_3$O$_6$ to PrBa$_2$Cu$_3$O$_6$.  The $x$-axis is
labelled with the $h$ or $l$ value of ${\bf q}$, where ${\bf q}$ =
($hh0$) or ($00l$), in reciprocal lattice units (rlu).  The
$\Sigma_1$ modes were measured at \mbox{${\bf Q}$ = ($3\!\pm\!h$,
$3\!\pm\!h$, 0)}, \mbox{($h$, $h$, 15)} and \mbox{($2\!-\!h$,
$2\!-\!h$, 0)}.  The $\Sigma_4$ modes were measured at \mbox{${\bf
Q}$ = ($h$, $h$, 7)} and \mbox{($h$, $h$, 8)}. The $\Lambda_1$ and
$\Lambda_5$ modes were measured at \mbox{${\bf Q}$ = (0, 0,
$7\!+\!l$)} and \mbox{(2, 2, $l$)} respectively.  Branches
$\Sigma_1$(1) and $\Sigma_1$(4) are plotted as open circles and
squares respectively.  Both are significantly shifted with respect
to the predictions of the model.  The open triangles in the
$\Lambda_1$ plot indicate modes that could not be indexed by
comparison with the model.  On IN22 there was no monitor to check
for accidental Bragg scattering, so it is possible that this was
the cause of these peaks.} \label{fig:LOandTOdispersion}
\end{center}
\end{figure}

To determine the dispersion of the observed phonon modes in an
unbiased way, each energy scan was fitted with a lineshape
constructed from several gaussians superimposed on a constant
background.  The centres, amplitudes and linewidths of the
gaussians and the height of the background were refined using a
least squares method.  Figure \ref{fig:LOandTOdispersion} shows
the dispersion curves obtained by plotting the peak centres as a
function of the phonon wavevector ${\bf q}$, where ${\bf q}$ is
obtained from

\begin{equation}
{\bf Q} = {\bf q} + \bm{\tau},
\end{equation}

\noindent and $\bm{\tau}$ is a reciprocal lattice vector.

\section{Model}

A model based on a common interatomic potential was previously
developed for the phonon dispersion curves in YBa$_2$Cu$_3$O$_6$.
\cite{Chaplot}  We adapted this for PrBa$_2$Cu$_3$O$_6$ by
inserting the appropriate lattice parameters, atomic masses and
nuclear scattering lengths, and the model was used to calculate
the frequencies and dynamical structure factors of the phonon
modes at the points in reciprocal space where our measurements
were performed.

Comparison of the ${\bf Q}$-variation of the predicted and
observed modes allowed us to identify the symmetries of the
different branches.  In this process the predicted frequencies
were used only as a guide.  More weight was given to the
correspondence between the predicted and observed intensities of
the modes.  To facilitate the comparison, simulations of the
energy-scans were produced by combining the predicted dynamical
structure factors with the spectrometer resolution function,
calculated using the approximate Cooper-Nathans method.
\cite{Cooper-Nathans}

Figs.\ \ref{fig:LOandTOscans} (e)--(h) show the simulations
corresponding to the selected raw data shown in plots (a)--(d).
Most of the observed modes were in good qualitative agreement with
the simulations, so assignment of the symmetries was clear.
However, plots (b) and (f) show a clear discrepancy between the
observed and predicted phonon spectra. We have assigned the large
peak observed at 33\,meV to the $\Sigma_1$(4) mode (for clarity we
have numbered the branches from highest to lowest in frequency).
Although the energy of this peak corresponds better to the
energies of the predicted $\Sigma_4(4)$ and $\Sigma_1(5)$ modes,
its intensity corresponds much better to that of the $\Sigma_1(4)$
mode, so we are confident that our assignation is correct.

\begin{figure}[!ht]
\begin{center}
\includegraphics{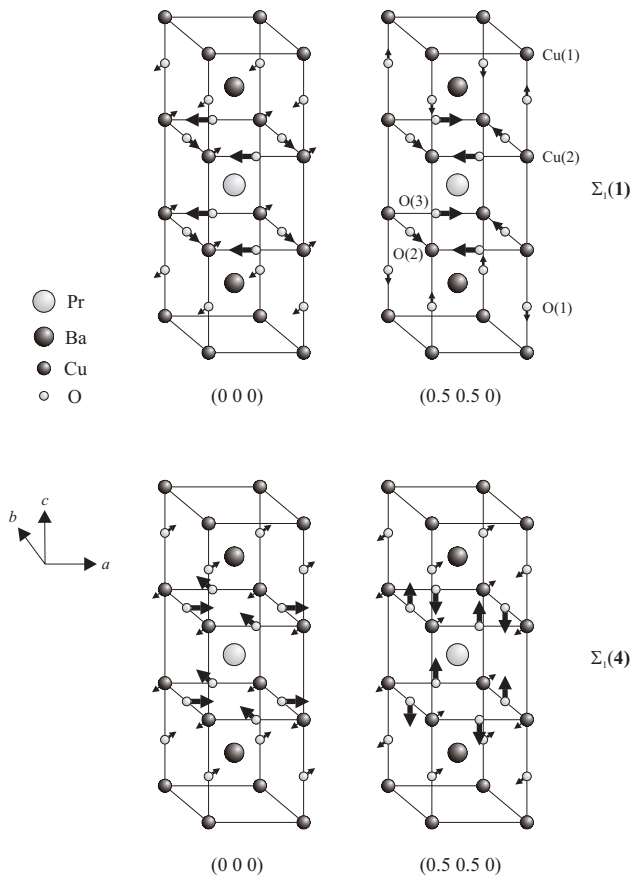}
\caption{Anomalous phonon modes.  Branches $\Sigma_1$(1) and
$\Sigma_1$(4) are shown at the Brillouin zone centre (0 0 0) and
edge (0.5 0.5 0). The length and thickness of the arrows are
approximately proportional to the magnitudes of the displacements
obtained from the common interatomic potential model for
PrBa$_2$Cu$_3$O$_{6+x}$ ($x = 0$).  The measured compound ($x
\approx 0.2$) has a partial occupancy of the two oxygen sites
halfway between the Cu sites in the basal plane.}
\label{fig:PhononModes}
\end{center}
\end{figure}

In Fig.\ \ref{fig:LOandTOdispersion} the measured dispersion
curves are compared with those predicted by the model.  The
agreement is good for the majority of the observed branches.
However, the model overestimates the energy of branch
$\Sigma_1$(1) by $\sim$ 8\,meV (2\,THz) at the zone centre and the
energy of branch $\Sigma_1$(4) by $\sim$ 10\,meV (2.5\,THz) over
the whole of the Brillouin zone.  The relative motions of the
atoms in these branches are shown in Fig.\ \ref{fig:PhononModes},
and it is clear that the motion of the copper and oxygen atoms in
the CuO$_2$ planes dominates.  Both branches change character
somewhat between the centre and the edge of the Brillouin zone, so
the diagrams indicate the motions of the atoms at both of these
positions in reciprocal space.

The broad resolution function of the spectrometer caused some
ambiguity in the assignation of mode symmetries at zone centre.
This was due to the possibility that we could have accidentally
detected out-of-plane modes propagating in transverse directions.
For instance, when we attempted to measure phonons propagating
along [hh0] it is possible that we could also have picked up modes
propagating along [00h].  This effect is only important at zone
centre, so although it could partially account for the lowering of
the $\Sigma_1(1)$ mode at zone centre, we are confident that it
could not have caused us to wrongly assign the symmetry of the
$\Sigma_1(4)$ mode.

We now compare the dispersion curves we have observed in
PrBa$_2$Cu$_3$O$_{6.2}$ with those previously observed in
YBa$_2$Cu$_3$O$_{6}$. \cite{Chaplot}  In YBa$_2$Cu$_3$O$_{6}$ all
the branches are well described by the common interatomic
potential model.  Therefore the energy shifts of the $\Sigma_1$(1)
and $\Sigma_1$(4) branches in PrBa$_2$Cu$_3$O$_{6.2}$ are
anomalous.  Although the $\Sigma_1$(1) branch in
YBa$_2$Cu$_3$O$_{6}$ has a similar shape to that in
PrBa$_2$Cu$_3$O$_{6.2}$ the dip at zone centre is not so
pronounced.  In contrast to PrBa$_2$Cu$_3$O$_{6.2}$, the
$\Sigma_1$(4) branch in YBa$_2$Cu$_3$O$_{6}$ fits well with the
model.  Although the energies of the other modes are very similar
in both compounds, the $\Sigma_1$(4) mode is $\sim$ 12\,meV
(3\,THz) higher in YBa$_2$Cu$_3$O$_{6}$ than in
PrBa$_2$Cu$_3$O$_{6.2}$ throughout the Brillouin zone.

The common interatomic potential model assumes that the binding
mechanism is predominantly ionic, but the anomalies we have
observed suggest the existence of more complex binding mechanisms.
It is particularly noteworthy that the anomalous phonon branches
are characterised by large displacements of the oxygen atoms whose
$2p$ orbitals are proposed to hybridise with the Pr $4f$ orbitals.
Pr--O hybridisation, in which the Pr--O bonds are partially
covalent in character, would result in a modified electron
distribution around the in-plane O atoms, which could change the
strength of the Cu--O bonds and alter certain vibrational modes.

\section{Conclusion}

We have described inelastic neutron scattering measurements of the
phonon dispersion curves in oxygen-deficient single crystal
PrBa$_2$Cu$_3$O$_{6+x}$ ($x \approx 0.2$).  The results have been
compared with a model of the lattice dynamics based on a common
interatomic potential, which is found to overestimate the
frequencies of two branches by a considerable amount. Strikingly,
these branches are dominated by motion of the oxygen atoms in the
CuO$_2$ planes: the same atoms that have been proposed to
hybridise with the Pr orbitals.  In contrast, the observed
frequencies of all phonon branches in YBa$_2$Cu$_3$O$_{6}$ agree
well with the model.  The frequency shifts observed in the two
anomalous branches in PrBa$_2$Cu$_3$O$_{6.2}$ are interpreted as
indirect evidence for hybridisation of the Pr 4$f$ and O 2$p$
orbitals.

\begin{acknowledgments}
We would like to thank L.-P. Regnault and A. Ivanov for help with
the experiments on IN22 and IN1 respectively.  Financial support
for C.H.G. by the EPSRC is also acknowledged.
\end{acknowledgments}

\bibliography{PrO2PrBCO}

\begin{thebibliography}{10}
\expandafter\ifx\csname natexlab\endcsname\relax\def\natexlab#1{#1}\fi
\expandafter\ifx\csname bibnamefont\endcsname\relax
  \def\bibnamefont#1{#1}\fi
\expandafter\ifx\csname bibfnamefont\endcsname\relax
  \def\bibfnamefont#1{#1}\fi
\expandafter\ifx\csname citenamefont\endcsname\relax
  \def\citenamefont#1{#1}\fi
\expandafter\ifx\csname url\endcsname\relax
  \def\url#1{\texttt{#1}}\fi
\expandafter\ifx\csname urlprefix\endcsname\relax\def\urlprefix{URL }\fi
\providecommand{\bibinfo}[2]{#2}
\providecommand{\eprint}[2][]{\url{#2}}

\bibitem[{\citenamefont{Radousky}(1992)}]{Radousky}
\bibinfo{author}{\bibfnamefont{H.~B.} \bibnamefont{Radousky}},
  \bibinfo{journal}{J. Mater. Res.} \textbf{\bibinfo{volume}{7}},
  \bibinfo{pages}{1917} (\bibinfo{year}{1992}).

\bibitem[{\citenamefont{Boothroyd}(2000)}]{Boothroyd:PrBCOReview}
\bibinfo{author}{\bibfnamefont{A.~T.} \bibnamefont{Boothroyd}},
  \bibinfo{journal}{J. Alloys Compd.} \textbf{\bibinfo{volume}{303--304}},
  \bibinfo{pages}{489} (\bibinfo{year}{2000}).

\bibitem[{\citenamefont{Fehrenbacher and Rice}(1993)}]{Fehrenbacher}
\bibinfo{author}{\bibfnamefont{R.}~\bibnamefont{Fehrenbacher}}
  \bibnamefont{and} \bibinfo{author}{\bibfnamefont{T.~M.} \bibnamefont{Rice}},
  \bibinfo{journal}{Phys. Rev. Lett.} \textbf{\bibinfo{volume}{70}},
  \bibinfo{pages}{3471} (\bibinfo{year}{1993}).

\bibitem[{\citenamefont{Lister et~al.}(2001)\citenamefont{Lister, Boothroyd,
  Andersen, Larsen, Zhokhov, Christensen, and Wildes}}]{Lister:2001}
\bibinfo{author}{\bibfnamefont{S.~J.~S.} \bibnamefont{Lister}},
  \bibinfo{author}{\bibfnamefont{A.~T.} \bibnamefont{Boothroyd}},
  \bibinfo{author}{\bibfnamefont{N.~H.} \bibnamefont{Andersen}},
  \bibinfo{author}{\bibfnamefont{B.~H.} \bibnamefont{Larsen}},
  \bibinfo{author}{\bibfnamefont{A.~A.} \bibnamefont{Zhokhov}},
  \bibinfo{author}{\bibfnamefont{A.~N.} \bibnamefont{Christensen}},
  \bibnamefont{and} \bibinfo{author}{\bibfnamefont{A.~R.}
  \bibnamefont{Wildes}}, \bibinfo{journal}{Phys. Rev. Lett.}
  \textbf{\bibinfo{volume}{86}}, \bibinfo{pages}{5994} (\bibinfo{year}{2001}).

\bibitem[{\citenamefont{Boothroyd et~al.}(1997)\citenamefont{Boothroyd,
  Longmore, Andersen, Brecht, and Wolf}}]{Boothroyd:1997}
\bibinfo{author}{\bibfnamefont{A.~T.} \bibnamefont{Boothroyd}},
  \bibinfo{author}{\bibfnamefont{A.}~\bibnamefont{Longmore}},
  \bibinfo{author}{\bibfnamefont{N.~H.} \bibnamefont{Andersen}},
  \bibinfo{author}{\bibfnamefont{E.}~\bibnamefont{Brecht}}, \bibnamefont{and}
  \bibinfo{author}{\bibfnamefont{T.}~\bibnamefont{Wolf}},
  \bibinfo{journal}{Phys. Rev. Lett.} \textbf{\bibinfo{volume}{78}},
  \bibinfo{pages}{130} (\bibinfo{year}{1997}).

\bibitem[{\citenamefont{Gardiner et~al.}(2002)\citenamefont{Gardiner, Lister,
  Boothroyd, Andersen, Zhokhov, Stunault, and Hiess}}]{Gardiner:2002:PrBCO}
\bibinfo{author}{\bibfnamefont{C.~H.} \bibnamefont{Gardiner}},
  \bibinfo{author}{\bibfnamefont{S.~J.~S.} \bibnamefont{Lister}},
  \bibinfo{author}{\bibfnamefont{A.~T.} \bibnamefont{Boothroyd}},
  \bibinfo{author}{\bibfnamefont{N.~H.} \bibnamefont{Andersen}},
  \bibinfo{author}{\bibfnamefont{A.~A.} \bibnamefont{Zhokhov}},
  \bibinfo{author}{\bibfnamefont{A.}~\bibnamefont{Stunault}}, \bibnamefont{and}
  \bibinfo{author}{\bibfnamefont{A.}~\bibnamefont{Hiess}},
  \bibinfo{journal}{Appl. Phys. A} \textbf{\bibinfo{volume}{74}},
  \bibinfo{pages}{S898} (\bibinfo{year}{2002}).

\bibitem[{\citenamefont{Chaplot et~al.}(1995)\citenamefont{Chaplot, Reichardt,
  Pintschovius, and Pyka}}]{Chaplot}
\bibinfo{author}{\bibfnamefont{S.~L.} \bibnamefont{Chaplot}},
  \bibinfo{author}{\bibfnamefont{W.}~\bibnamefont{Reichardt}},
  \bibinfo{author}{\bibfnamefont{L.}~\bibnamefont{Pintschovius}},
  \bibnamefont{and} \bibinfo{author}{\bibfnamefont{N.}~\bibnamefont{Pyka}},
  \bibinfo{journal}{Phys. Rev. B} \textbf{\bibinfo{volume}{52}},
  \bibinfo{pages}{7230} (\bibinfo{year}{1995}).

\bibitem[{\citenamefont{Squires}(1996)}]{Squires}
\bibinfo{author}{\bibfnamefont{G.~L.} \bibnamefont{Squires}},
  \emph{\bibinfo{title}{Introduction to the Theory of Thermal Neutron
  Scattering}} (\bibinfo{publisher}{Dover Publications, Inc.},
  \bibinfo{address}{Mineola, New York, USA}, \bibinfo{year}{1996}).

\bibitem[{\citenamefont{Hilscher et~al.}(1994)\citenamefont{Hilscher,
  Holland-Moritz, Holubar, Jostarndt, Nekvasil, Schaudy, Walter, and
  Fillion}}]{Hilscher}
\bibinfo{author}{\bibfnamefont{G.}~\bibnamefont{Hilscher}},
  \bibinfo{author}{\bibfnamefont{E.}~\bibnamefont{Holland-Moritz}},
  \bibinfo{author}{\bibfnamefont{T.}~\bibnamefont{Holubar}},
  \bibinfo{author}{\bibfnamefont{H.-D.} \bibnamefont{Jostarndt}},
  \bibinfo{author}{\bibfnamefont{V.}~\bibnamefont{Nekvasil}},
  \bibinfo{author}{\bibfnamefont{G.}~\bibnamefont{Schaudy}},
  \bibinfo{author}{\bibfnamefont{U.}~\bibnamefont{Walter}}, \bibnamefont{and}
  \bibinfo{author}{\bibfnamefont{G.}~\bibnamefont{Fillion}},
  \bibinfo{journal}{Phys. Rev. B} \textbf{\bibinfo{volume}{49}},
  \bibinfo{pages}{535} (\bibinfo{year}{1994}).

\bibitem[{\citenamefont{Cooper and Nathans}(1967)}]{Cooper-Nathans}
\bibinfo{author}{\bibfnamefont{M.~J.} \bibnamefont{Cooper}} \bibnamefont{and}
  \bibinfo{author}{\bibfnamefont{R.}~\bibnamefont{Nathans}},
  \bibinfo{journal}{Acta. Cryst.} \textbf{\bibinfo{volume}{23}},
  \bibinfo{pages}{357} (\bibinfo{year}{1967}).

\end{thebibliography}

\end{document}